\documentclass{ws-ijmpa}

\makeatletter
\let\trimmarks\relax
\makeatother

\usepackage[super]{cite}
\usepackage{xcolor}
\usepackage[verbose,hypertexnames=false]{hyperref}
\hypersetup{colorlinks=false,allbordercolors=blue,pdfborderstyle={/S/U/W 1}}

\makeatletter
\def\@clinebuf{}
\makeatother

\begin{document}

\markboth{Emese Árpási, László Kovács, Dániel Kincses, Máté Csanád}{Source shapes in coordinate and momentum space in $\sqrt{s_\text{NN}}$ = 200 GeV Au+Au collisions}

%
%

\title{Source shapes in coordinate and momentum space in $\sqrt{s_\text{NN}}$ = 200 GeV Au+Au collisions}

\author{Emese Árpási, László Kovács, Dániel Kincses, Máté Csanád}

\address{Department of Atomic Physics, ELTE Eötvös Loránd University\\Pázmány Péter sétány 1/A, Budapest, H-1117, Hungary\\
earpasi@student.elte.hu, laszlokov@staff.elte.hu, kincses@ttk.elte.hu, csanad@elte.hu}

\maketitle
\pagestyle{plain}

\begin{abstract}
In high-energy heavy-ion physics, femtoscopy aims to explore and understand the space-time characteristics of the collisions through investigating the correlation function of particle pairs. In recent years, experimental measurements have shown that the shape of the particle source function can be well described by the Lévy distribution, motivating further phenomenological studies. In this work, we perform a three-dimensional analysis of the source shapes for different particle-pair types in $\sqrt{s_\text{NN}}=200$ GeV Au+Au collisions generated with the EPOS event generator. In contrast to our previous analyses, where we fitted the one-dimensional projections of the pair distribution function, $D(\rho_j)$ ($j \in \{\text{out}, \text{side}, \text{long}\}$), now we reconstruct the three-dimensional $C(\boldsymbol{q})$ correlation function and fit it directly, as in experimental analyses. Furthermore, we now apply full three-dimensional fits instead of fitting the one-dimensional projections.  
We find that this approach results in smaller Lévy index values, while the direction averaged Lévy scale and the correlation strength remain largely unchanged, except for kaons. In the case of pions, the $\alpha$ values are closer to the PHENIX data, showing a good agreement above 20\% centrality.
\end{abstract}

\section{Introduction}
In high-energy heavy-ion physics, femtoscopy is an experimental method that provides a tool for probing the space-time structure of processes occurring after the collisions on the femtometer scale [\protect\refcite{Lisa:2005dd}]. In recent years, phenomenological studies have gained increasing importance alongside experimental investigations in femtoscopy. In the present work, we perform a phenomenological analysis of two-particle correlation functions of same-charged identical particle pairs (namely pions, kaons and protons), using collision events generated with the EPOS event generator.

Over the past decades, increasingly precise experimental measurements have highlighted the limitations of describing particle-emitting source functions using a Gaussian [\protect\refcite{PHENIX:2007grx,Brown:2000aj}]. This has motivated investigations beyond the Gaussian approximation and extending the functional form of the source function to a more general form, among which the Lévy distribution provides a useful description [\protect\refcite{Csorgo:2003uv,Csanad:2024hva}]. Several experimental analyses have already been performed using this approach [\protect\refcite{PHENIX:2024vjp,Kincses:2024sin,NA61SHINE:2023qzr,CMS:2023xyd,PHENIX:2017ino}], providing support for the applicability of Lévy-shaped source functions. Alongside these experimental studies, several phenomenological investigations have been initiated to explore the properties and physical implications of Lévy-shaped source functions [\protect\refcite{Kincses:2026tfa,Kincses:2025izu,Kincses:2024lnv,Csanad:2024jpy,Korodi:2022ohn,Kincses:2019rug,Kincses:2025iaf}].

The present study builds on our previous investigations [\protect\refcite{Kincses:2026tfa,Kincses:2025izu}]. We compare the results obtained using two different analysis methods: fitting the pair-distribution function and fitting the correlation function directly. We investigate how the difference between the methods affects the obtained parameters. Our results show that the choice of fitting method mostly affects the Lévy index, while the Lévy scale and the correlation strength remain mostly unchanged. We also compare our new pion results with the previous findings in Ref. [\protect\refcite{Kincses:2025izu}] and the PHENIX measurement [\protect\refcite{PHENIX:2024vjp}]. The structure of the article is the following: in Section 2, we introduce the new analysis method, in Section 3, we show our results and compare them with the previous results, and finally in Section 4 we conclude.

\section{Analysis methods}
In this study, we analysed Au+Au collisions at $\sqrt{s_\text{{NN}}}=200 \hspace{3pt} \mathrm{GeV}$ center-of-mass energy, generated using EPOS [\protect\refcite{Werner:2010aa,Werner:2007bf,Werner:2013tya}] (version 359). The analysis was performed on a sample of 300,000 events. We investigated the correlation function of identically charged pion, kaon and proton pairs.

In order to determine the correlation function, we first reconstructed the spatial pair distribution function of the produced particles from their freeze-out coordinates. For each identical-particle pairs, we calculated their relative spatial distance vector, and transformed it into the Longitudinally Co-Moving System (LCMS) [\protect\refcite{Csorgo:1991ej}]. During the calculation of the pair distribution, we expressed the relative distance in the Bertsch--Pratt coordinate system [\protect\refcite{Pratt:1986ev,Bertsch:1988db}], where the long direction is parallel to the beam axis, the out direction is parallel to the average transverse momentum of the particle pair and side is perpendicular to the other two.

In contrast to the previous analyses [\protect\refcite{Kincses:2026tfa,Kincses:2025izu}], in which we reconstructed the pair distribution function $D(\boldsymbol{\rho})$ and fitted its one-dimensional projections $D(\rho_j)$ ($j \in \{\text{out}, \text{side}, \text{long}\}$), in the present study, we fitted the $C(\boldsymbol{q})$ correlation function derived from the particle distributions obtained from the simulated events. As a first step, we reconstructed the three-dimensional pair source function $D(\boldsymbol{\rho})$, from which the correlation function can be calculated as [\protect\refcite{Koonin:1977fh,Pratt:1984su,Goldhaber:1960sf,YANO1978556}]: 

\begin{align}
 C(\boldsymbol{q}) = \int D(\boldsymbol{\rho}) |\Psi_{\boldsymbol{q}}(\boldsymbol{\rho})|^2 d^3\boldsymbol{\rho}
 \label{e:cq}
\end{align}
Here $\boldsymbol{\rho}$ denotes the relative distance vector, $\boldsymbol{q}$ is the relative momentum vector of the particle pair and $\Psi_{\boldsymbol{q}}(\boldsymbol{\rho})$ is the pair wave function.

Without final-state interactions, using the plane-wave approximation for the two-particle wave function, the correlation function for identical bosons can be expressed as
\begin{gather}
    C(\boldsymbol{q}) = 1 + \int D(\boldsymbol{\rho}) \cos\left(\frac{\boldsymbol{q\boldsymbol{\rho}}}{\hbar c}\right) d^3\boldsymbol{\rho} = 1 + \operatorname{Re}[\widetilde{D}(\boldsymbol{q})], \label{eq:Cq} \\
    \text{where } \tilde{D}(\boldsymbol{q}) = \int D(\boldsymbol{\rho}) e^{i \boldsymbol{q} \boldsymbol{\rho} / \hbar c} \, \mathrm{d}^3 \boldsymbol{\rho}.
\end{gather}

To reconstruct $D(\boldsymbol{\rho})$, we used the same method and cuts as mentioned in articles [\protect\refcite{Kincses:2026tfa,Kincses:2025izu}], with the difference that we reconstructed the full three-dimensional distribution function rather than only its three one-dimensional projections. The $D(\boldsymbol{\rho})$ distribution was reconstructed in the range of $[-100, 100]\text{ fm}$ along each spatial direction. The reconstructed distribution was normalized by its total integral, including the underflow and overflow bins. Subsequently, we calculated its Fourier transform, using the Fast Fourier Transform method (FFT):

\begin{align}
    \widetilde{D}(\boldsymbol{q_m}) = \sum_{\boldsymbol{n}=0}^{N-1} D(\boldsymbol{\rho_n})\exp\left(i\frac{2\pi}{N}\boldsymbol{m}\boldsymbol{n}\right),
\end{align}
where $\boldsymbol{m} = (m_{\mathrm{out}}, m_{\mathrm{side}}, m_{\mathrm{long}})$ and $\boldsymbol{n} = (n_{\mathrm{out}}, n_{\mathrm{side}}, n_{\mathrm{long}}) \in \{0, \dots, N-1\}^3$ denote the grid indices in momentum and coordinate space, respectively. The momentum grid resolution is determined as follows:

\begin{align}
    \Delta q_j = \dfrac{2\pi}{N \Delta\rho_j} \hbar c, \quad j \in \{\text{out}, \text{side}, \text{long\}},
\end{align}
with $\hbar c$ = 0.197327 GeV$\cdot$fm.
After the transformation, the correlation function can be obtained as given in \eqref{eq:Cq}.
We used the resulting three-dimensional correlation functions in the following fitting procedure.

We divided the generated events into centrality classes, to investigate the centrality dependence of the obtained parameters. For the particle-species dependence analysis, we considered three centrality classes: 0-10\%, 10-20\% and 20-30\%. For the centrality dependence analysis of pions, we used six centrality classes in the 0-60\% centrality range. Furthermore, within each centrality class, we divided the particle pairs into transverse mass bins, defined as $m_\text{T}=\sqrt{m_\text{{particle}}^2+k_\text{T}^2}$, where $m_{\text{particle}}$ is the mass of the given particle, and $k_\text{T}$ is the average transverse momentum of the pair. This binning allowed us to investigate the transverse mass dependence of the obtained parameters. The reconstruction of the correlation function was carried out within these centrality and transverse mass bins.

\subsection{Fitting method}
We parametrised the source function using a three-dimensional Lévy distribution, from which the pair distribution function takes the form as:
\begin{align}
    D(\boldsymbol{\rho})=\lambda\mathcal{L}(\boldsymbol{\rho},\alpha,R) {.}
\end{align}
Here, $\lambda$ is the correlation strength, and $\mathcal{L}$ denotes the three-dimensional Lévy function, defined as:
\begin{align}
    \mathcal{L}(\boldsymbol{\rho},\alpha,R)=\frac{1}{(2\pi)^3}\int\mathrm{d}^3\boldsymbol{\xi} e^{i\boldsymbol{\xi \rho}} e^{-\left|\boldsymbol{\xi}^T\boldsymbol{R}^2\boldsymbol{\xi}\right|^{\alpha/2}},
\end{align}
where the $\boldsymbol{R}^2$ matrix contains the squared Lévy scale parameters, which characterize the geometric sizes in the corresponding directions. The parameter $\alpha$ is the Lévy index, characterizing the shape, and $\boldsymbol{\xi}$ is a three-dimensional integration variable.

The correlation function was fitted using the Fourier transform of the Lévy shaped pair distribution function. For the correlation function, we fit the following three-dimensional function $C(\boldsymbol{q})$:
\begin{align}
    C(\boldsymbol{q}) = 1 + \lambda \mathrm{exp}\left(-\left| \frac{\boldsymbol{q}^T\boldsymbol{R}^2\boldsymbol{q}}{(\hbar c)^2}\right|^{\alpha/2}\right). \label{eq:fit}
\end{align}

For protons, Eq.~\eqref{eq:Cq} and the corresponding fit function in Eq.~\eqref{eq:fit} are applied consistently both to generate and to fit the momentum-space distribution as a Bose-like mathematical transform of $D(\boldsymbol{\rho})$. The physical proton–proton correlation function without final-state interactions, $C(\boldsymbol{q}) = 1 - \frac{1}{2}\operatorname{Re}[\widetilde{D}(\boldsymbol{q})]$, would yield exactly the same source parameters.

Since the Fast Fourier Transform of the reconstructed $D(\boldsymbol{\rho})$ does not yield direct statistical bin errors for $C(\boldsymbol{q})$, in the initial fit uniform uncertainties were assigned to the bins, which were then rescaled so that $\chi^2/\mathrm{NDF} = 1.00$ for the best fit, allowing an appropriate error estimation for the parameters.

\begin{figure*}[h]
    \centering
    \includegraphics[trim=0 20bp 0 0, clip, width=0.9\linewidth]{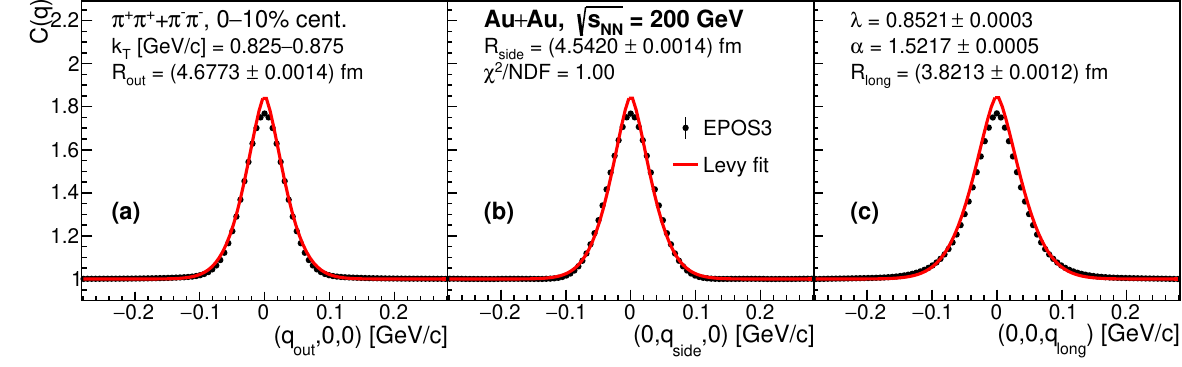}\\[-1ex]
    \vspace{-2.5mm}
    \includegraphics[trim=0 20bp 0 0, clip, width=0.9\linewidth]{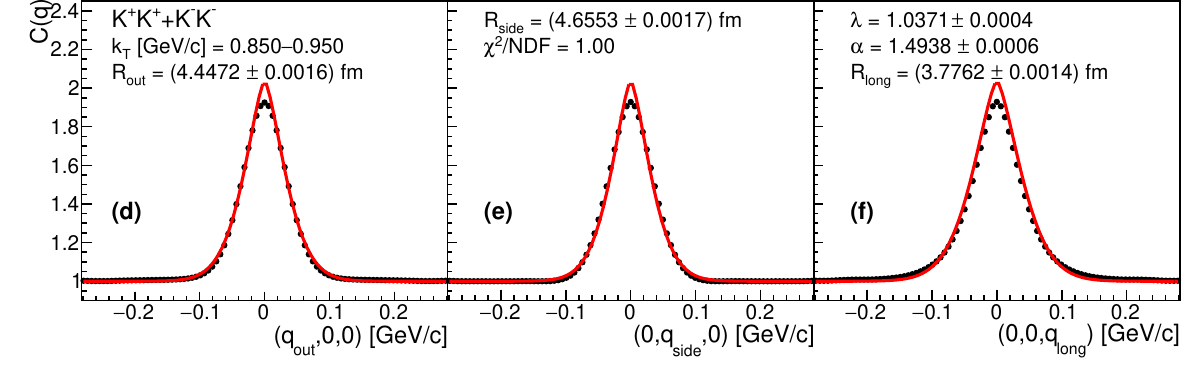}\\[-1ex]
    \vspace{-2.5mm}
    \includegraphics[trim=0 0 0 0, clip, width=0.9\linewidth]{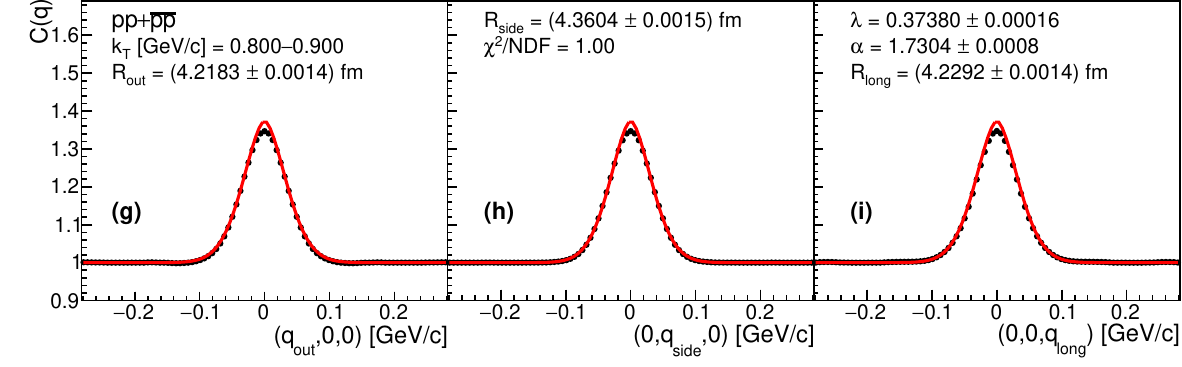}
    
    \caption{Example fits for the three different particle species for centrality 0-10\%. The different rows show the different particle species (pions, kaons and protons), while the columns show one-dimensional slices along the out, side, and long axes.}
    \label{fig:fit}
\end{figure*}

Figure \ref{fig:fit} shows example fits to the $C(\boldsymbol{q})$ correlation function for the three particle types we studied for the 0-10\% centrality class. In the different columns there are the results for the three slices, while the rows correspond to the different particle species.
\subsection{Systematic uncertainties}
To estimate the systematic uncertainties, we considered three major sources. We investigated the impact of the choice of the number of histogram bins ($N_\text{bins}$), the upper limit of the fit range ($q^\text{fit}_\text{max}$), and the maximum pair relative momentum ($Q_\text{LCMS}^\text{max}$), applied as in Ref.~[\protect\refcite{Kincses:2026tfa,Kincses:2025izu}]. The fits were performed within a spherical momentum range defined by 
\begin{equation}
    \sqrt{q_{\text{out}}^2 + q_{\text{side}}^2 + q_{\text{long}}^2} \le q_\text{max}^\text{fit}.
\end{equation} For each parameter, we defined a default value, then we varied it to evaluate the effect on the extracted results.

The values used in the systematic studies were the following:
\begin{itemize}
    \item $N_\text{bins}$: 211, (111, 311)
    \item $q_\text{max}^\text{fit}$: 0.3, (0.2, 0.4) [GeV/c]
    \item $Q_\text{LCMS}^\text{max}=\sqrt{A\cdot m_\text{T}}$:
    \begin{itemize}
        \item[--] pions: $A = 0.15$ ($0.05$, $0.25$) [$\mathrm{GeV}$]
        \item[--] kaons: $A = 0.15$ ($0.05$, $0.25$) [$\mathrm{GeV}$]
        \item[--] protons: $A = 0.15$ ($0.10$, $0.20$) [$\mathrm{GeV}$]
\end{itemize}
\end{itemize}
Our final total systematic uncertainty was calculated as the quadratic sum of all the individual systematic uncertainties. Among the three sources investigated, variation of the $Q^\text{max}_\text{LCMS}$ cut had the largest impact on the obtained parameters.

Note that here, unlike in our previous studies [\protect\refcite{Kincses:2026tfa,Kincses:2025izu}], we did not investigate the dependence on the number of events averaged ($N_\text{event}$), but we averaged over all the events within one centrality class.

\section{Results and discussion}
As described in Section 2, from the fitting we obtained the five parameters of the correlation function ($\alpha$, $R_\text{out},R_\text{side},R_\text{long}$, $\lambda$). In this section, we compare these parameters with the results of our previous studies [\protect\refcite{Kincses:2026tfa,Kincses:2025izu}]. We first investigate their dependence on the particle species and the transverse mass by comparing the results obtained from fits to the source function  $D(\rho_j)$ with those obtained from fits to the correlation function $C(\boldsymbol{q})$. We then examine the centrality dependence and compare our results with the PHENIX measurement [\protect\refcite{PHENIX:2024vjp}].

\subsection{Particle-species dependence}
\begin{figure*}[h]
    \centering
    \includegraphics[width=\linewidth]{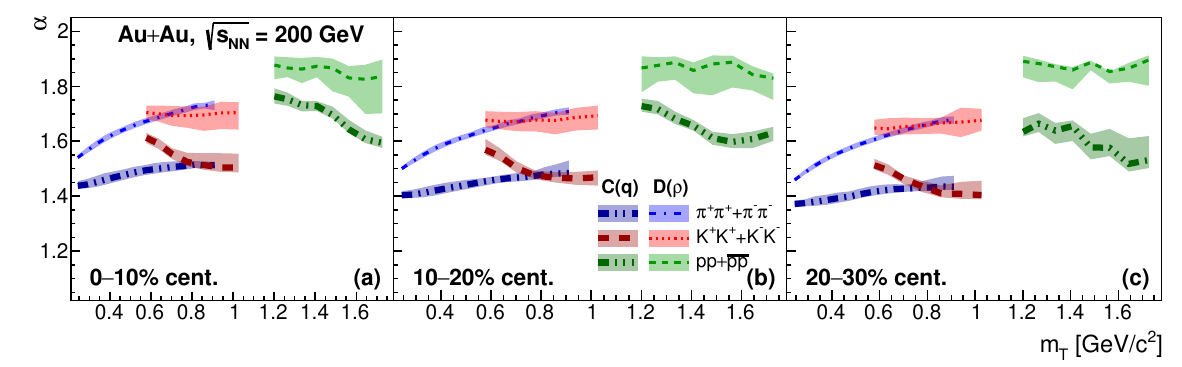}
    \caption{Lévy exponent parameter $\alpha$ as a function of average transverse mass $m_\text{T}$. The three centrality classes are shown in separate panels, while on a given panel, the values corresponding to the three different particle species are shown with different colours and line styles. The bands around the dashed lines correspond to the total systematic uncertainty. The $C(\boldsymbol{q})$ results are the topic of this paper, while the $D(\rho_j)$ results are from Ref. ~\protect\refcite{Kincses:2026tfa}.}
    \label{fig:alpha}
\end{figure*}

In Figure \ref{fig:alpha}, we can see the transverse mass ($m_\text{T}$) dependence of the Lévy index $\alpha$ for the three particle species, in three different centrality classes. The values obtained from fits to the $C(\boldsymbol{q})$ correlation function are lower than those obtained from fits to $D(\rho_j)$, for all particle types, for all centrality classes. Furthermore, in our previous analysis based on fits to $D(\rho_j)$, the $\alpha$ values showed a continuous increasing tendency with $m_\text{T}$ across the different particle species. This trend is not observed in the results obtained from fits to $C(\boldsymbol{q})$, indicating a difference in the observed transverse-mass dependence between the two methods.  

\begin{figure*}
    \centering
    \includegraphics[trim=0 20bp 0 0, clip, width=0.9\linewidth]{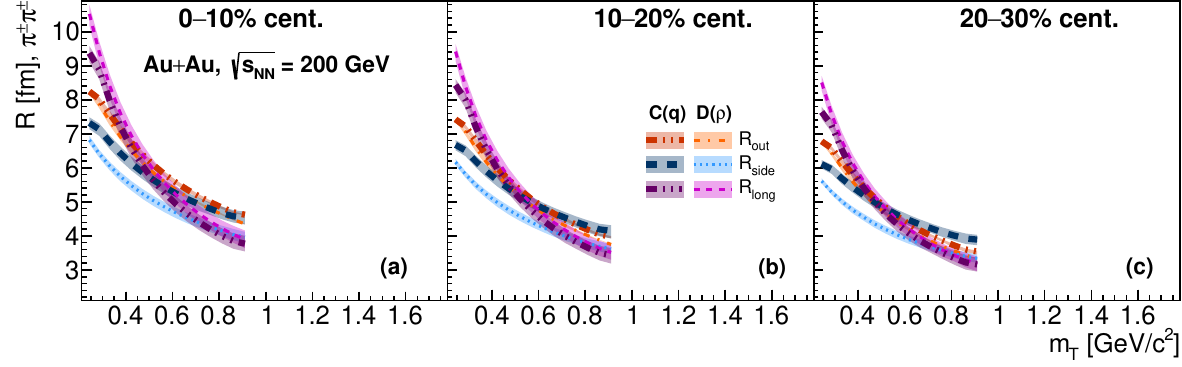}\\[-1ex]
    \vspace{-2.5mm}
    \includegraphics[trim=0 20bp 0 0, clip, width=0.9\linewidth]{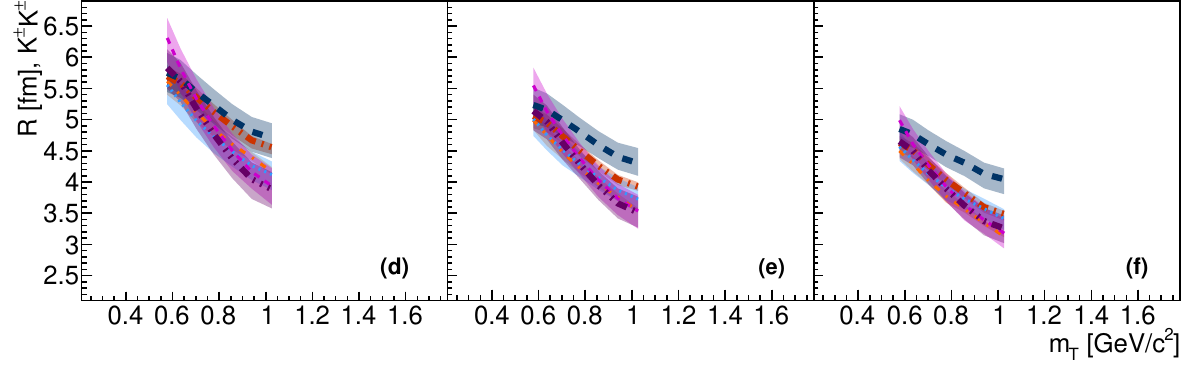}\\[-1ex]
    \vspace{-2.5mm}
    \includegraphics[trim=0 0 0 0, clip, width=0.9\linewidth]{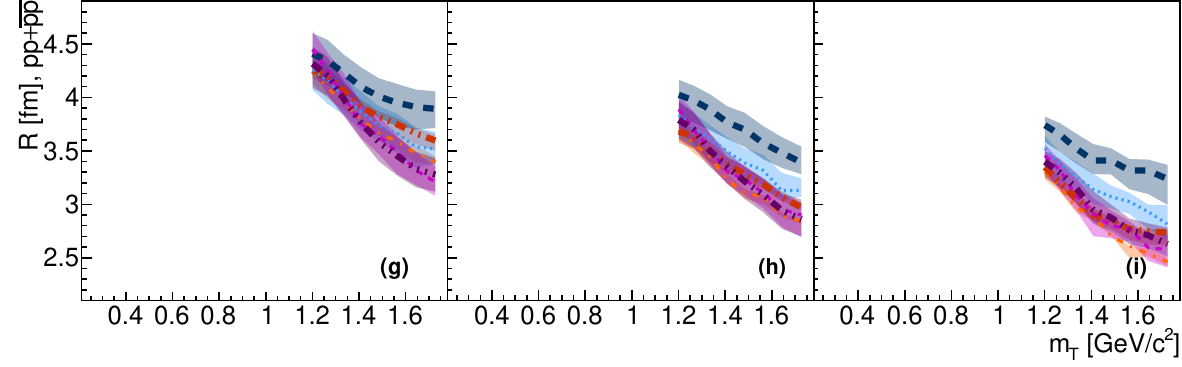}
    
    \caption{Lévy scale parameters $R_{\rm out,side,long}$ as a function of average transverse mass $m_\text{T}$. The three centrality classes are shown in separate columns, and the three particle species are shown in separate rows. On a given panel, the values corresponding to the three different directions (out,side,long) are shown with different colours and line styles. The bands around the dashed lines correspond to the total systematic uncertainty. The $C(\boldsymbol{q})$ results are the topic of this paper, while the $D(\rho_j)$ results are from Ref. ~\protect\refcite{Kincses:2026tfa}.}
    \label{fig:R3D}
\end{figure*}

For the three Lévy scale parameters $R_\text{out},R_\text{side}$, and $R_\text{long}$, the obtained values compared to the previous results are in Figure \ref{fig:R3D}. The results from the $D(\rho_j)$ fitting and the results from the $C(\boldsymbol{q})$ fitting are consistent for the out and long directions for all particle species and all centrality classes. In contrast, $R_\text{side}$ values obtained using the new Fourier-transform-based method are larger than those obtained from the pair distribution function fits.

\begin{figure*}
    \centering
    \includegraphics[width=\linewidth]{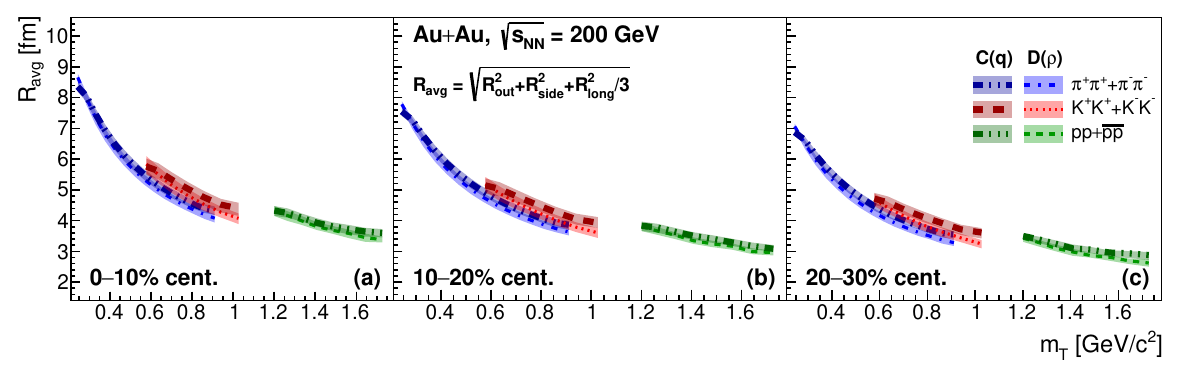}
    \caption{Average Lévy scale parameter $R_{\rm avg}$ as a function of average transverse mass $m_\text{T}$. The three centrality classes are shown in separate panels, while on a given panel, the values corresponding to the three different particle species are shown with different colours and line styles. The bands around the dashed lines correspond to the total systematic uncertainty. The $C(\boldsymbol{q})$ results are the topic of this paper, while the $D(\rho_j)$ results are from Ref. ~\protect\refcite{Kincses:2026tfa}.}
    \label{fig:Ravg}
\end{figure*}

In Figure \ref{fig:Ravg}, we plotted the averaged Lévy radii parameters $R_\text{avg}=\sqrt{(R_\text{out}^2+R_\text{side}^2+R_\text{long}^2)/3}$ as a function of transverse mass for the three particle species, and for the three centrality classes. Despite the differences in the individual radii, the average radii obtained using the two methods are in good agreement, and the tendencies are similar.

\begin{figure*}
    \centering
    \includegraphics[width=\linewidth]{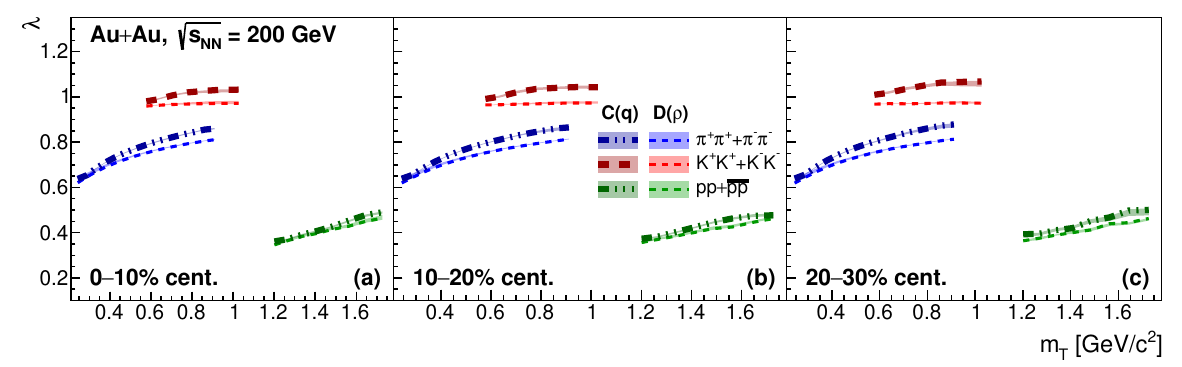}
    \caption{Source strength parameter $\lambda$ as a function of average transverse mass $m_\text{T}$. The three centrality classes are shown in separate panels, while on a given panel, the values corresponding to the three different particle species are shown with different colours and line styles. The bands around the dashed lines correspond to the total systematic uncertainty. The $C(\boldsymbol{q})$ results are the topic of this paper, while the $D(\rho_j)$ results are from Ref. ~\protect\refcite{Kincses:2026tfa}, where the normalization cutoff $\rho_{\text{max}}^\lambda = 5\text{ cm}$ was applied as detailed in Ref. ~\protect\refcite{Kincses:2025izu}.} 
    \label{fig:lambda}
\end{figure*}

In Figure \ref{fig:lambda}, we plotted the obtained correlation strength values as a function of the transverse mass with the results from our previous study. The two methods resulted in good agreements for this parameter for protons, showed a minor difference for pions for larger $m_\text{T}$ values and resulted in a larger disagreement for kaons.

For the latter, $\lambda$ slightly exceeds unity. The reconstructed source is, to a good approximation, a three-dimensional elliptical Lévy distribution, but its shape may differ between directions and at large distances, which a single Lévy index and three scale parameters cannot capture simultaneously. The fit therefore compromises, and since the few bins near $\boldsymbol{q}$ = 0 carry little weight in a three-dimensional fit, the extrapolated intercept $1 + \lambda$ ends up above the reconstructed $C(\boldsymbol{q} = 0)$ (see Figure \ref{fig:lambda}). As nearly all same-charge kaon pairs originate from the core, this pushes $\lambda$ above unity only for kaons, while for pions and protons the effect is masked by the halo contribution.

\subsection{Centrality dependence}
\begin{figure*}[h]
    \centering
    \includegraphics[width=0.95\linewidth]{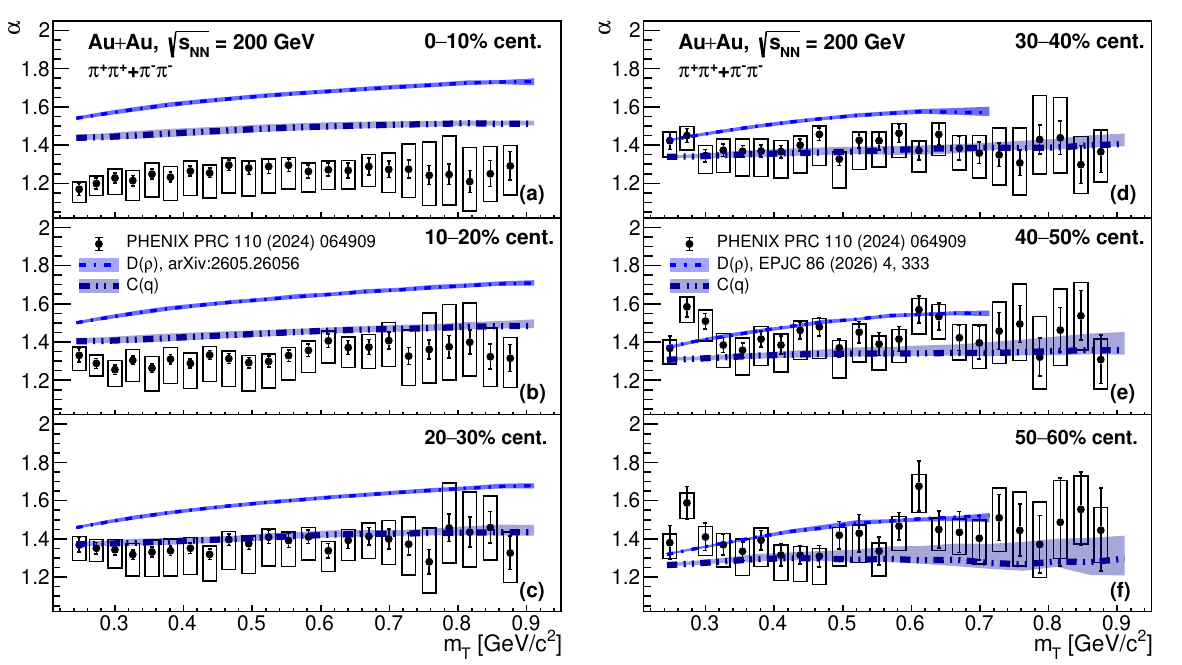}
    \caption{Lévy index parameter $\alpha$ as a function of averaged transverse mass $m_\text{T}$, compared to the PHENIX measurement [\protect\refcite{PHENIX:2024vjp}]. The six centrality classes are shown in separate panels, while on a given panel, the values corresponding to our result are shown with coloured lines, different line styles for the $C(\boldsymbol{q})$ and $D(\rho_j)$ methods. The bands around the dashed lines correspond to the total systematic uncertainty. The black dots with black boxes denote the PHENIX data values. The $C(\boldsymbol{q})$ results are the topic of this paper, while the $D(\rho_j)$ results are from Ref.  ~\protect\refcite{Kincses:2026tfa} in the case of centralities 0-30\% and Ref. ~\protect\refcite{Kincses:2025izu} in the case of centralities 30-60\%.}
    \label{fig:alpha_phenix}
\end{figure*}

Figure \ref{fig:alpha_phenix} shows the obtained $\alpha$ values for the six centrality classes for the two methods, compared to the PHENIX results from Ref. [\protect\refcite{PHENIX:2024vjp}]. The obtained $\alpha$ values from the $C(\boldsymbol{q})$ fits are in a much better agreement with the experimental results, than the $\alpha$ values obtained from the $D(\rho_j)$ fits. Our new results are in a good agreement with the experimental data above 20\% centrality, while for our previous results, the agreement is only above 40-50\%.

\begin{figure*}[h]
    \centering
    \includegraphics[width=0.7\linewidth]{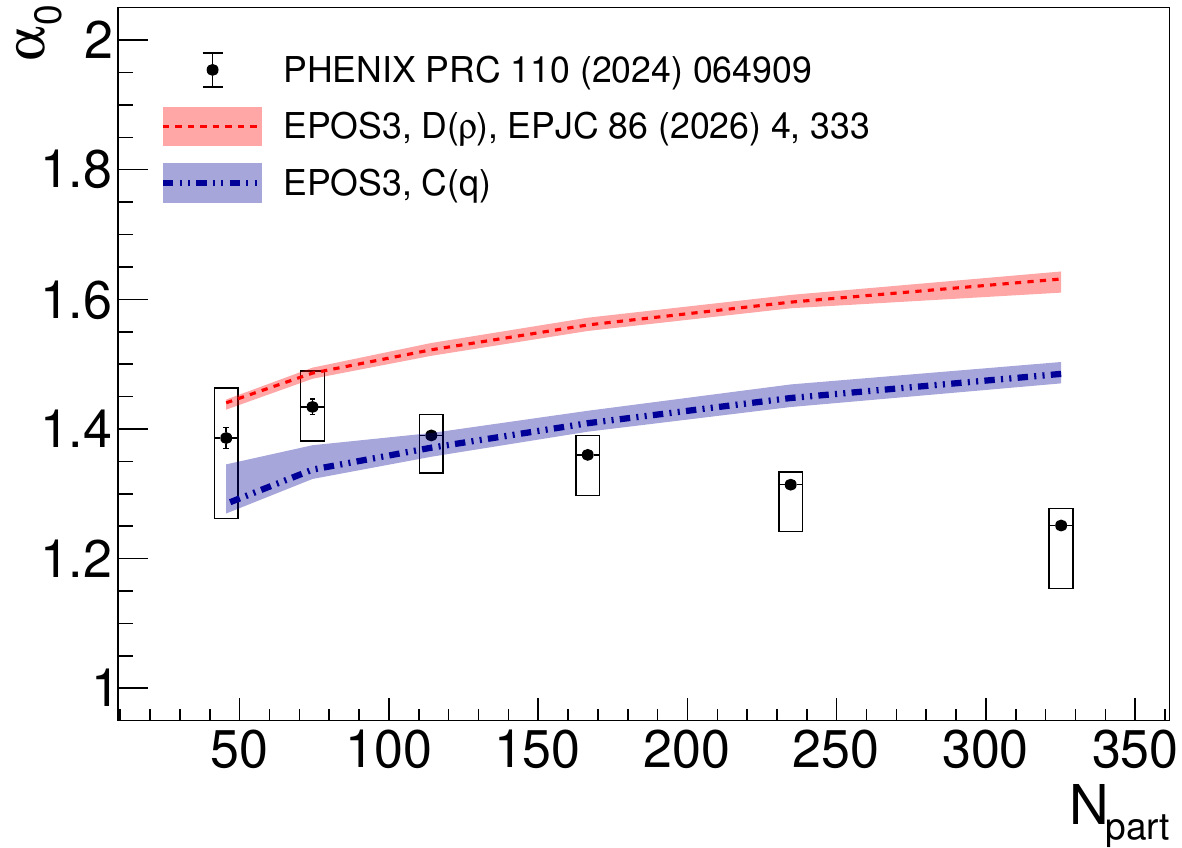}
    \caption{The $m_\text{T}$-averaged Lévy index parameter $\alpha_0$ as a function of the number of participants, compared to the PHENIX measurement [\protect\refcite{PHENIX:2024vjp}]. The values corresponding to our result are shown with coloured lines, the blue line denotes the $C(\boldsymbol{q})$ and the red line denotes $D(\rho_j)$ methods. The bands around the dashed lines correspond to the total systematic uncertainty. The black dots with black boxes denote the PHENIX data values. The $C(\boldsymbol{q})$ results are the topic of this paper, while the $D(\rho_j)$ results are from Ref. ~\protect\refcite{Kincses:2025izu}.}
    \label{fig:alpha0}
\end{figure*}

Figure \ref{fig:alpha0} shows the $m_\text{T}$-averaged Lévy index values, $\alpha_0 = \langle \alpha_{m_\text{T}} \rangle$, as a function of the number of participants ($N_{\textrm{part}}$). We compared our new results with the $D(\rho_j)$ fits and the PHENIX data from Ref.[\protect\refcite{PHENIX:2024vjp}]. While our new results are closer to the experimental data, they still show a large difference for central events, and the trend is opposite to the experimentally measured one, same as obtained from the $D(\rho_j)$ fits.

\begin{figure*}[h]
    \centering
    \includegraphics[width=0.95\linewidth]{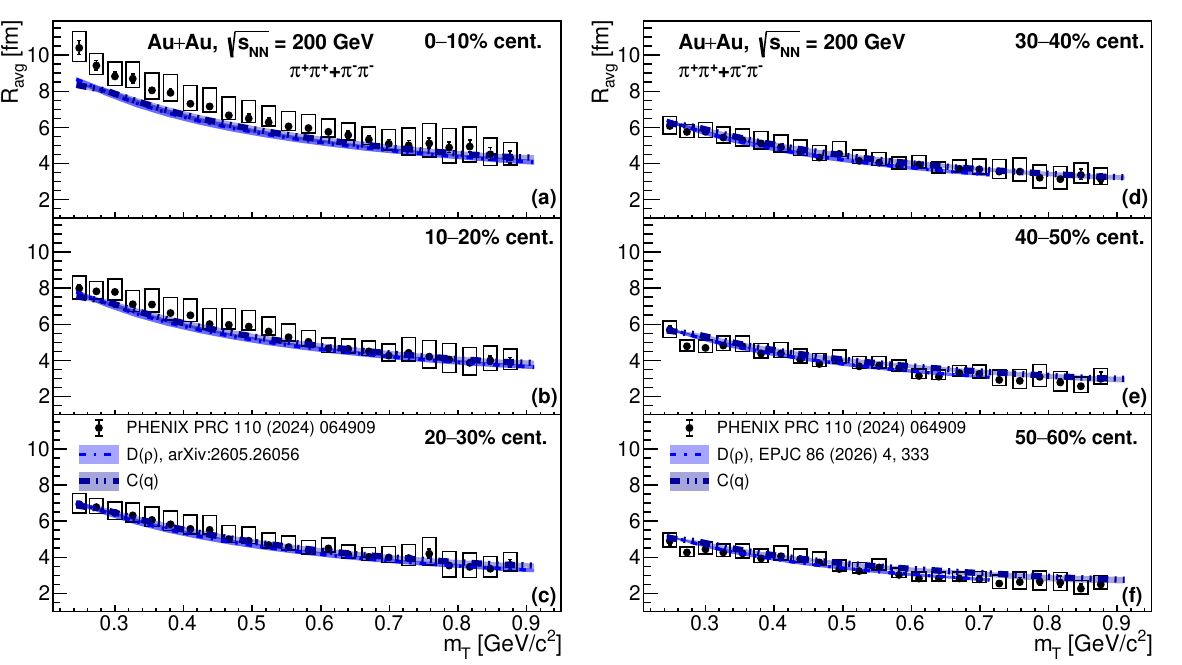}
    \caption{Direction averaged Lévy radii parameter $R_{avg}$ as a function of averaged transverse mass $m_\text{T}$, compared to the PHENIX measurement [\protect\refcite{PHENIX:2024vjp}]. The six centrality classes are shown in separate panels, while on a given panel, the values corresponding to our result are shown with coloured lines, different line styles for the $C(\boldsymbol{q})$ and $D(\rho_j)$ methods. The bands around the dashed lines correspond to the total systematic uncertainty. The black dots with black boxes denote the PHENIX data values. The $C(\boldsymbol{q})$ results are the topic of this paper, while the $D(\rho_j)$ results are from Ref.  ~\protect\refcite{Kincses:2026tfa} in the case of centralities 0-30\% and Ref. ~\protect\refcite{Kincses:2025izu} in the case of centralities 30-60\%.}
    \label{fig:ravg_phenix}
\end{figure*}

Figure \ref{fig:ravg_phenix} shows the direction averaged Lévy radii parameters' ($R_\text{avg}$) dependence on the averaged transverse mass for charged pion pairs, for six centrality classes compared to the PHENIX data. The values are in a good agreement with the experimental results above 10\%. The simulation also captured the tendencies well.

\begin{figure*}[h]
    \centering
    \includegraphics[width=0.95\linewidth]{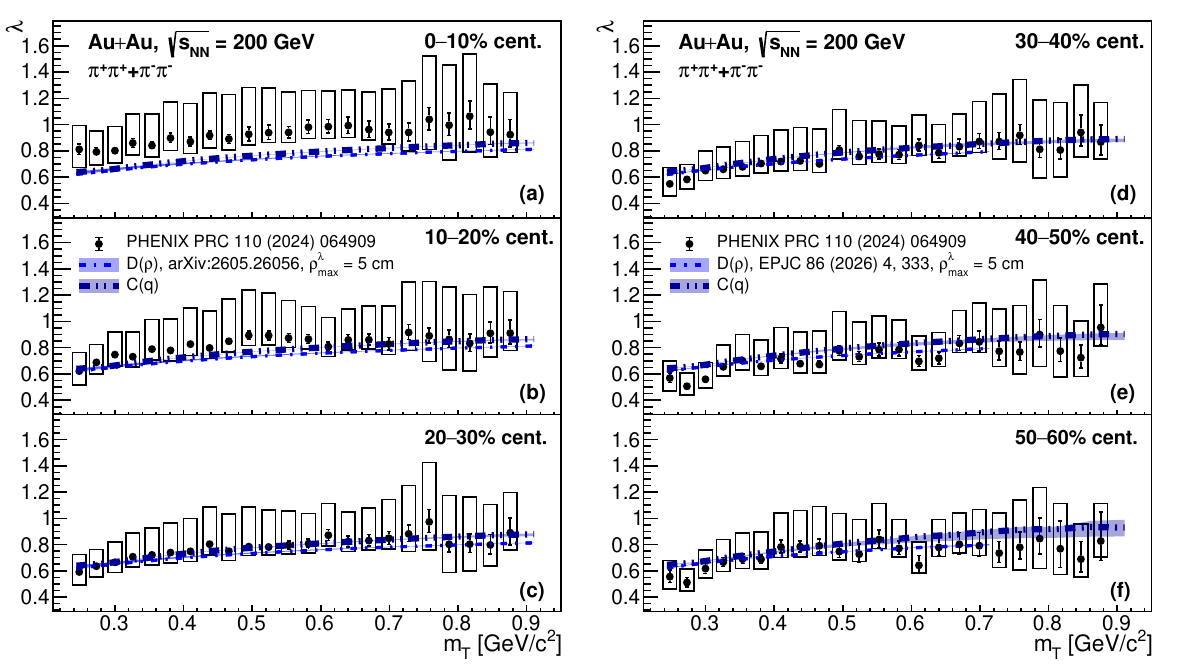}
    \caption{Correlation strength parameter $\lambda$ as a function of averaged transverse mass $m_\text{T}$, compared to the PHENIX measurement [\protect\refcite{PHENIX:2024vjp}]. The six centrality classes are shown in separate panels, while on a given panel, the values corresponding to our result are shown with coloured lines, different line styles for the $C(\boldsymbol{q})$ and $D(\rho_j)$ methods. The bands around the dashed lines correspond to the total systematic uncertainty. The black dots with black boxes denote the PHENIX data values. The $C(\boldsymbol{q})$ results are the topic of this paper, while the $D(\rho_j)$ results are taken from Ref.~\protect\refcite{Kincses:2026tfa} for 0–30\% and Ref.~\protect\refcite{Kincses:2025izu} for 30–60\% centralities; in both cases, the normalization cutoff $\rho_{\text{max}}^\lambda = 5\text{ cm}$ was applied as detailed in Ref. ~\protect\refcite{Kincses:2025izu}.}
    \label{fig:lambda_phenix}
\end{figure*}

Figure \ref{fig:lambda_phenix} shows the correlation strength ($\lambda$) values as a function of the averaged transverse mass ($m_\text{T}$) for the two analysis methods, for six centrality classes, compared to the PHENIX data. The magnitude and the trends were well captured by EPOS, the results from the simulation showed a good agreement above 10\% with the experimental data.

The comparison with PHENIX data is affected by the choice of fitting method only for the pion Lévy index $\alpha$, as the scale and correlation strength yield consistent results.

\section{Summary}

In our study, we investigated the shape of the correlation function of identical particle pairs, using $\sqrt{s_\text{NN}}=200 \hspace{3 pt} \mathrm{GeV}$ Au+Au events generated with the EPOS event generator. We used the Lévy distribution as the source function, and we fitted the $C(\boldsymbol{q})$ correlation function reconstructed from the events. We also compared the obtained parameters ($\alpha$, $R_\text{out},R_\text{side},R_\text{long}$, $\lambda$) with our previous results from articles [\protect\refcite{Kincses:2026tfa,Kincses:2025izu}], where instead of the $C(\boldsymbol{q})$ we reconstructed and fitted the pair distribution $D(\rho_j)$ ($j \in \{\text{out}, \text{side}, \text{long}\}$). Furthermore, now we fitted the full three-dimensional functions, rather than the one-dimensional projections. We investigated the particle-species dependence of the parameters using charged pions, kaons, and protons, and compared our pion results with the PHENIX results from article [\protect\refcite{PHENIX:2024vjp}].

In the case of the particle species study, we saw that from the five parameters, the largest difference between the two methods occurred in the case of the Lévy index ($\alpha$). For every centrality class and particle species, we found that the $\alpha$ values obtained with the $C(\boldsymbol{q})$ fits are smaller than the ones obtained from the previous method. For the Lévy scale values in the three directions we got similar values as in our previous studies, the largest difference was at the side direction, for the $R_\text{side}$ values, where we obtained larger values than for the $D(\rho_j)$ fits. When it comes to the direction averaged Lévy scale value ($R_\text{avg}$), we obtained an agreement within the systematic uncertainties. For the correlation strength, we only observed a significant difference for kaons, in the case of pions and protons we found a good agreement between the two methods.

For the centrality dependence study, we analysed the correlation function of charged pion pairs for six centrality classes from 0-60\%, and compared the results with the PHENIX data [\protect\refcite{PHENIX:2024vjp}]. For the Lévy index parameter ($\alpha$), we found that regarding the $m_\text{T}$ dependence, while the trends are captured well by the simulation, the magnitudes differ for the most central cases, and agree for the more peripheral cases. However, the new $C(\boldsymbol{q})$ fitting method resulted in a better agreement with the experimental data, than the previous $D(\rho_j)$ fitting method, now we obtained agreement above centrality 20\%. For the $m_\text{T}$-averaged Lévy index values ($\alpha_0$) versus the number of participants ($N_{\textrm{part}}$), we found opposite trends between our results and the PHENIX data, although the new results from $C(\boldsymbol{q})$ fits are closer to the experimental results. For the direction averaged Lévy scale ($R_{\textrm{avg}}$) and correlation strength ($\lambda$) values, we obtained similar results for the $D(\rho_j)$ fits and the $C(\boldsymbol{q})$ fits, our results are in a good agreement with the PHENIX data above centrality 10\%.

Our results show that the analysis method influences the extracted parameters, especially the Lévy index, while the Lévy scale and the correlation strength remain almost unchanged. Furthermore, fitting to the correlation function instead of the pair distribution function produces a better agreement with the experimental data. This outcome indicates the need for further investigation to find the source of this discrepancy.

\section*{Acknowledgement}
E.Á. and L.K. were supported by the EKÖP-26 University Research Scholarship 
Program of the Ministry of Science and Technology from the source of the National
Research, Development and Innovation Fund. Further support was provided by NKFIH under grant numbers 2026-4.1.1-MISSZIÓK-2026-00005, PD-146589, and NKKP ADVANCED 152097.

\clearpage
\bibliographystyle{ws-ijmpa}
\bibliography{bibliography}

@article{Kincses:2025izu,
    author = "Kincses, Daniel and Árpási, Emese and Kovács, Lászlo and Nagy, Marton and Csanád, Mate",
    title = "{Three-dimensional sizes and shapes of pion emission in heavy-ion collisions}",
    eprint = "2511.01808",
    archivePrefix = "arXiv",
    primaryClass = "nucl-ex",
    doi = "10.1140/epjc/s10052-026-15526-1",
    journal = "Eur. Phys. J. C",
    volume = "86",
    number = "4",
    pages = "333",
    year = "2026"
}

@article{PHENIX:2024vjp,
    author = "Abdulameer, N. J. and others",
    collaboration = "PHENIX",
    title = "{Centrality dependence of L{\'e}vy-stable two-pion Bose-Einstein correlations in sNN=200 GeV Au+Au collisions}",
    eprint = "2407.08586",
    archivePrefix = "arXiv",
    primaryClass = "nucl-ex",
    doi = "10.1103/PhysRevC.110.064909",
    journal = "Phys. Rev. C",
    volume = "110",
    number = "6",
    pages = "064909",
    year = "2024"
}

@article{Kincses:2026tfa,
    author = "Kincses, D{\'a}niel and Kov{\'a}cs, L{\'a}szl{\'o} and Csan{\'a}d, M{\'a}t{\'e}",
    title = "{Particle species dependence of femtoscopic source parameters in high-energy nuclear collisions}",
    eprint = "2605.26056",
    archivePrefix = "arXiv",
    primaryClass = "hep-ph",
    month = "",
    year = "2026"
}

@article{Lisa:2005dd,
    author = "Lisa, Michael Annan and Pratt, Scott and Soltz, Ron and Wiedemann, Urs",
    title = "{Femtoscopy in relativistic heavy ion collisions}",
    eprint = "nucl-ex/0505014",
    archivePrefix = "arXiv",
    doi = "10.1146/annurev.nucl.55.090704.151533",
    journal = "Ann. Rev. Nucl. Part. Sci.",
    volume = "55",
    pages = "357--402",
    year = "2005"
}

@article{PHENIX:2007grx,
    author = "Afanasiev, S. and others",
    collaboration = "PHENIX",
    title = "{Source breakup dynamics in Au+Au Collisions at s(NN)**(1/2) = 200-GeV via three-dimensional two-pion source imaging}",
    eprint = "0712.4372",
    archivePrefix = "arXiv",
    primaryClass = "nucl-ex",
    doi = "10.1103/PhysRevLett.100.232301",
    journal = "Phys. Rev. Lett.",
    volume = "100",
    pages = "232301",
    year = "2008"
}

@article{Brown:2000aj,
    author = "Brown, D. A. and Danielewicz, P.",
    title = "{Observing nongaussian sources in heavy ion reactions}",
    eprint = "nucl-th/0010108",
    archivePrefix = "arXiv",
    reportNumber = "DOE-ER-41132-100-INT00",
    doi = "10.1103/PhysRevC.64.014902",
    journal = "Phys. Rev. C",
    volume = "64",
    pages = "014902",
    year = "2001"
}

@article{Csorgo:2003uv,
    author = "Csörgő, T. and Hegyi, S. and Zajc, W. A.",
    title = "{Bose-Einstein correlations for Levy stable source distributions}",
    eprint = "nucl-th/0310042",
    archivePrefix = "arXiv",
    doi = "10.1140/epjc/s2004-01870-9",
    journal = "Eur. Phys. J. C",
    volume = "36",
    pages = "67--78",
    year = "2004"
}

@article{Csanad:2024hva,
    author = "Csan{\'a}d, M{\'a}t{\'e} and Kincses, D{\'a}niel",
    title = "{Femtoscopy with L{\'e}vy Sources from SPS through RHIC to LHC}",
    eprint = "2401.01249",
    archivePrefix = "arXiv",
    primaryClass = "hep-ph",
    doi = "10.3390/universe10020054",
    journal = "Universe",
    volume = "10",
    number = "2",
    pages = "54",
    year = "2024"
}

@article{Kincses:2024sin,
    author = "Kincses, D{\'a}niel",
    collaboration = "STAR",
    title = "{Pion Interferometry with L{\'e}vy-Stable Sources in = 200 GeV Au + Au Collisions at STAR}",
    eprint = "2401.11169",
    archivePrefix = "arXiv",
    primaryClass = "nucl-ex",
    doi = "10.3390/universe10030102",
    journal = "Universe",
    volume = "10",
    number = "3",
    pages = "102",
    year = "2024"
}

@article{NA61SHINE:2023qzr,
    author = "Adhikary, H. and others",
    collaboration = "NA61/SHINE",
    title = "{Two-pion femtoscopic correlations in Be+Be collisions at $\sqrt{s_{\text {NN}}} = 16.84$ GeV measured by the NA61/SHINE~at CERN}",
    eprint = "2302.04593",
    archivePrefix = "arXiv",
    primaryClass = "nucl-ex",
    reportNumber = "FERMILAB-PUB-23-057-AD",
    doi = "10.1140/epjc/s10052-023-11997-8",
    journal = "Eur. Phys. J. C",
    volume = "83",
    number = "10",
    pages = "919",
    year = "2023"
}

@article{CMS:2023xyd,
    author = "Tumasyan, Armen and others",
    collaboration = "CMS",
    title = "{Two-particle Bose-Einstein correlations and their L{\'e}vy parameters in PbPb collisions at sNN=5.02 TeV}",
    eprint = "2306.11574",
    archivePrefix = "arXiv",
    primaryClass = "nucl-ex",
    reportNumber = "CMS-HIN-21-011, CERN-EP-2023-094",
    doi = "10.1103/PhysRevC.109.024914",
    journal = "Phys. Rev. C",
    volume = "109",
    number = "2",
    pages = "024914",
    year = "2024"
}

@article{Kincses:2024lnv,
    author = "Kincses, D{\'a}niel and Nagy, M{\'a}rton and Csan{\'a}d, M{\'a}t{\'e}",
    title = "{L{\'e}vy walk of pions in heavy-ion collisions}",
    eprint = "2409.10373",
    archivePrefix = "arXiv",
    primaryClass = "nucl-th",
    doi = "10.1038/s42005-025-01973-x",
    journal = "Commun. Phys.",
    volume = "8",
    number = "1",
    pages = "55",
    year = "2025"
}

@article{Csanad:2024jpy,
    author = "Csanád, Mate and Kincses, Daniel",
    title = "{Investigating the excitation function of HBT radii for L{\'e}vy-stable sources}",
    eprint = "2406.11435",
    archivePrefix = "arXiv",
    primaryClass = "nucl-th",
    doi = "10.1088/1361-6471/ada04f",
    journal = "J. Phys. G",
    volume = "52",
    number = "2",
    pages = "025102",
    year = "2025"
}

@article{Korodi:2022ohn,
    author = "K{\'o}rodi, Bal{\'a}zs and Kincses, D{\'a}niel and Csan{\'a}d, M{\'a}t{\'e}",
    title = "{Event-by-event investigation of the two-particle source function in sNN=2.76 TeV PbPb collisions with EPOS}",
    eprint = "2212.02980",
    archivePrefix = "arXiv",
    primaryClass = "nucl-th",
    doi = "10.1016/j.physletb.2023.138295",
    journal = "Phys. Lett. B",
    volume = "847",
    pages = "138295",
    year = "2023"
}

@article{PHENIX:2017ino,
    author = "Adare, A. and others",
    collaboration = "PHENIX",
    title = "{L{\'e}vy-stable two-pion Bose-Einstein correlations in $\sqrt{s_{NN}}=200$ GeV Au$+$Au collisions}",
    eprint = "1709.05649",
    archivePrefix = "arXiv",
    primaryClass = "nucl-ex",
    doi = "10.1103/PhysRevC.97.064911",
    journal = "Phys. Rev. C",
    volume = "97",
    number = "6",
    pages = "064911",
    year = "2018",
}

@article{Kincses:2019rug,
    author = "Kincses, D. and Nagy, M. I. and Csan{\'a}d, M.",
    title = "{Coulomb and strong interactions in the final state of Hanbury-Brown{\textendash}Twiss correlations for L{\'e}vy-type source functions}",
    eprint = "1912.01381",
    archivePrefix = "arXiv",
    primaryClass = "hep-ph",
    doi = "10.1103/PhysRevC.102.064912",
    journal = "Phys. Rev. C",
    volume = "102",
    number = "6",
    pages = "064912",
    year = "2020"
}

@article{Kincses:2025iaf,
    author = "Kincses, Daniel",
    title = "{Femtoscopic signatures of unique nuclear structures in relativistic collisions}",
    eprint = "2506.05849",
    archivePrefix = "arXiv",
    primaryClass = "nucl-th",
    doi = "10.1103/1y5m-j966",
    journal = "Phys. Rev. Res.",
    volume = "7",
    number = "4",
    pages = "L042028",
    year = "2025"
}

@article{Werner:2010aa,
    author = "Werner, K. and Karpenko, Iu. and Pierog, T. and Bleicher, M. and Mikhailov, K.",
    title = "{Event-by-Event Simulation of the Three-Dimensional Hydrodynamic Evolution from Flux Tube Initial Conditions in Ultrarelativistic Heavy Ion Collisions}",
    eprint = "1004.0805",
    archivePrefix = "arXiv",
    primaryClass = "nucl-th",
    doi = "10.1103/PhysRevC.82.044904",
    journal = "Phys. Rev. C",
    volume = "82",
    pages = "044904",
    year = "2010"
}

@article{Werner:2007bf,
    author = "Werner, Klaus",
    title = "{Core-corona separation in ultra-relativistic heavy ion collisions}",
    eprint = "0704.1270",
    archivePrefix = "arXiv",
    primaryClass = "nucl-th",
    doi = "10.1103/PhysRevLett.98.152301",
    journal = "Phys. Rev. Lett.",
    volume = "98",
    pages = "152301",
    year = "2007"
}

@article{Werner:2013tya,
    author = "Werner, K. and Guiot, B. and Karpenko, Iu. and Pierog, T.",
    title = "{Analysing radial flow features in p-Pb and p-p collisions at several TeV by studying identified particle production in EPOS3}",
    eprint = "1312.1233",
    archivePrefix = "arXiv",
    primaryClass = "nucl-th",
    doi = "10.1103/PhysRevC.89.064903",
    journal = "Phys. Rev. C",
    volume = "89",
    number = "6",
    pages = "064903",
    year = "2014"
}

@article{Pratt:1986ev,
    author = "Pratt, S.",
    title = "{Coherence and Coulomb Effects on Pion Interferometry}",
    doi = "10.1103/PhysRevD.33.72",
    journal = "Phys. Rev. D",
    volume = "33",
    pages = "72--79",
    year = "1986"
}

@article{Bertsch:1988db,
    author = "Bertsch, G. and Gong, M. and Tohyama, M.",
    title = "{Pion Interferometry in Ultrarelativistic Heavy Ion Collisions}",
    reportNumber = "MSUCL-639",
    doi = "10.1103/PhysRevC.37.1896",
    journal = "Phys. Rev. C",
    volume = "37",
    pages = "1896--1900",
    year = "1988"
}

@article{Csorgo:1991ej,
    author = "{T. Cs\"org\H{o} and S. Pratt}",
    title = "{Structure of the peak in Bose-Einstein correlations}",
    journal = "Proceedings of the Relativistic Heavy Ion Physics at Present and Future Accelerators",
    volume = "KFKI-1991-28/A",
    pages = "75--90",
    year = "1991"
}

@article{Goldhaber:1960sf,
    author = "Goldhaber, Gerson and Goldhaber, Sulamith and Lee, Won-Yong and Pais, Abraham",
    editor = "Weiner, R. M.",
    title = "{Influence of Bose-Einstein statistics on the anti-proton proton annihilation process}",
    doi = "10.1103/PhysRev.120.300",
    journal = "Phys. Rev.",
    volume = "120",
    pages = "300--312",
    year = "1960"
}

@article{YANO1978556,
title = {Determining pion source parameters in relativistic heavy-ion collisions},
journal = {Phys. Lett. B},
volume = {78},
number = {5},
pages = {556-559},
year = {1978},
issn = {0370-2693},
doi = "10.1016/0370-2693(78)90638-X",
url = {https://www.sciencedirect.com/science/article/pii/037026937890638X},
author = {F.B. Yano and S.E. Koonin}
}

@article{Koonin:1977fh,
    author = "Koonin, S. E.",
    title = "{Proton Pictures of High-Energy Nuclear Collisions}",
    doi = "10.1016/0370-2693(77)90340-9",
    journal = "Phys. Lett. B",
    volume = "70",
    pages = "43--47",
    year = "1977"
}

@article{Pratt:1984su,
    author = "Pratt, S.",
    editor = "Weiner, R. M.",
    title = "{Pion Interferometry for Exploding Sources}",
    doi = "10.1103/PhysRevLett.53.1219",
    journal = "Phys. Rev. Lett.",
    volume = "53",
    pages = "1219--1221",
    year = "1984"
}

\end{document}